\documentclass[aps,prb,twocolumn,a4paper,10pt]{revtex4-1} 

\usepackage[T1]{fontenc}		
\usepackage[english]{babel}		
\usepackage[latin1]{inputenc}	
\usepackage{times}

\usepackage{amsmath}			
\usepackage{amssymb}			
\usepackage{bbm}				

\usepackage[colorlinks=true, a4paper=true, pdfstartview=FitV,linkcolor=blue, citecolor=blue, urlcolor=blue]{hyperref}

\usepackage{graphicx}			
\usepackage{graphics}			
\DeclareGraphicsExtensions{.pdf}
\usepackage{color}		

\newcommand{\bea}{\begin{eqnarray}}
\newcommand{\eea}{\end{eqnarray}}

\begin{document}
\title{Spontaneous Breakdown of Time-Reversal Symmetry Induced by Thermal Fluctuations}

\author{Johan~Carlstr\"om~and~Egor~Babaev
}
\affiliation{ 
${}^1$ Department of Theoretical Physics, The Royal Institute of Technology, Stockholm, SE-10691 Sweden  
}

\date{\today}

\begin{abstract}
In systems with broken $U(1)$ symmetry, such as superfluids, superconductors, or magnets, the symmetry restoration is driven by the proliferation of topological defects in the form of vortex loops (unless the phase transition is strongly first order). Here we discuss that   the proliferation of topological defects can, by contrast, lead to the breakdown of an additional symmetry. 
We demonstrate that this effect  should take place in  $s+is$ superconductors, which are widely discussed in connection with the iron-based materials (although the mechanism is much more general). In these systems a vortex excitation can create a "bubble" of fluctuating $Z_2$ order parameter. The thermal excitation of vortices then leads to breakdown of $Z_2$ time-reversal symmetry when the temperature 
is  {\it increased}. 
\end{abstract}
\maketitle

Usually 
states which break symmetries and exhibit long- or quasi-long-range order (such as superconductors, superfluids and ordered magnetic states) 
form at low temperatures. 
For example, three-dimensional conventional superconductors and superfluids break $U(1)$ local and global symmetries respectively \footnote{Arguments have been advanced recently that superfluidity does not necessary require broken $U(1)$ symmetry even in three dimensions but can also arise from $U(1)$-like degeneracies in the free energy \cite{prl14}}.
At elevated  temperatures, fluctuations destroy the order and symmetry is restored. 
The rather generic mechanism that drives this phase transition in superfluids is proliferation of vortex loops that 
 destroy  long-range order 
in the corresponding order parameter field $ | \psi(\mathbf{r})|e^{i\varphi (\mathbf{r})} $ \cite{Onsager,DH,peskin}  (unless the system has a strong first order 
phase transition like type-I superconductors \cite{coleman,hlm}).  

Similarly, in two-dimensional superfluids, the transition to the normal state is driven by proliferation of topological defects in the form of unbinding of vortex-antivortex pairs
\cite{KT}. 

Likewise, in systems with different symmetries,
phase transitions to more symmetric states are driven by proliferation of corresponding topological defects. Examples of this include domain walls in systems that break $Z_2$ symmetry, or bound states of topological defects in systems with multiple broken symmetries\cite{nature,kuklov,herland,berg,agterberg}.

In this work we demonstrate that topological defects can play a radically different role in certain systems, and instead
lead to spontaneous breakdown of a symmetry which is {\it not} broken in the ground state.
We specifically focus on frustrated three-band superconductors, but the scenario is by no means limited to this case.

This effect arises when fluctuations are included in the multi-component Ginsburg-Landau free energy density that describes the $s+is$ superconducting state:
\bea\nonumber
H= \sum_{a=1}^3 \Big\{\frac{1}{2}|\mathbf{D}\psi_a|^2 +\alpha_a |\psi_a|^2+\frac{\beta_a}{2} |\psi_a|^4\Big\}\\
+\frac{1}{2}(\nabla\times \mathbf{A})^2+\sum_{a\not=b}\eta_{ab}|\psi_a||\psi_b|\cos(\varphi_a-\varphi_b),\label{Hamiltonian}
\eea
where $\mathbf{D}=\nabla+ie\mathbf{A}$ is the covariant derivative and  $\psi_a=|\psi_a| e^{i\varphi_a}$ are complex fields representing, for example the superconducting components in different bands. The last terms in (\ref{Hamiltonian}) represent Josephson-Leggett interband
coupling. The magnetic field is given by $\mathbf{B}=\nabla\times \mathbf{A}$.
The $s+is$ state is realized due to frustration with respect to the phase differences between superconducting components. 
This for instance occurs if all $\eta_{ab}$ are positive since the last terms in Eq. \ref{Hamiltonian} then are minimised by all phase differences being $\varphi_a-\varphi_b=\pi$, which cannot be simultaneously satisfied. Likewise, the case with one positive and two negative couplings is also frustrated. 
Such a situation also occurs in systems with more than three components \cite{weston}. 
The  $s+is$ s  state is  currently attracting substantial
interest in connection with iron-based  \cite{maiti,nagaosa,tesanovic} 
as well as other kinds of superconductors \cite{PhysRevB.84.134520}.
Inter-component interactions of this type can also be realized in 
cold atoms experiments.

\begin{figure}[!htb]
\includegraphics[width=\linewidth]{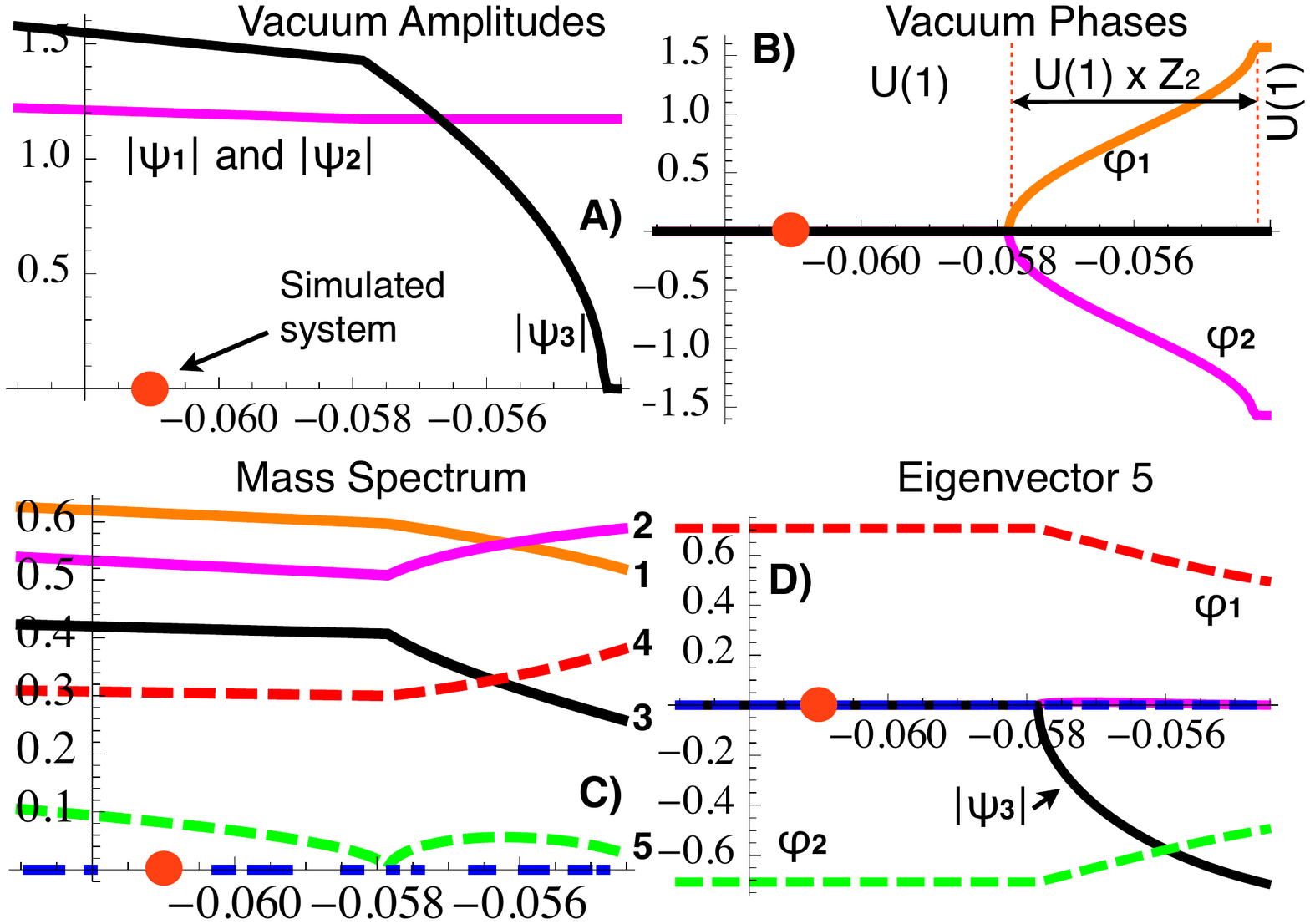}
\caption{
Ground state and mass spectrum of frustrated three-band GL model. Two of the model parameters, $\eta_{13}=\eta_{23}$ are scaled on the x-axis. The others are given by
$\alpha_1=\alpha_2=-3/64$ , $\alpha_3=2.673/64$, 
$\beta_1=\beta_2=3/64$, $\beta_3=0.18/64$, $\eta_{12}=2.25/64$ and $e=1.8/8. $
(In this example the parameters were choosen such that the relevant length scales of the theory to be
larger than numerical lattice spacing and substantially smaller than the system size.)
{The panel (A)} gives the ground state amplitudes of the fields while {(B)} gives the phases ($\varphi_3$ is fixed to $0$).
A critical point at $\eta_{13}=\eta_{23}\approx -0.0578$ separates 
a region with broken $U(1)$ symmetry (left) from a region that also breaks time-reversal symmetry (right). At a second critical point $\eta_{13}=\eta_{23}\approx -0.0542$, time-reversal symmetry is restored. The broken symmetries are indicated in (B).
(C) gives the mass spectrum and thus the the inverse of the coherence lengths (in units of lattice spacings) associated with the three fields. For detailed discussion of definition of coherence lengths/mass spectrum in this kind of models see \cite{Johan2011}. One mass (5) becomes zero at the critical point, implying a diverging coherence length. The mode that corresponds to this coherence length is shown in (D). On the left side it describes a perturbation to the phases $\varphi_1,\varphi_2$, i.e. a Leggett mode, which becomes massless at the critical point.  In the region to the right of the critical point the character of the mode changes, as it now describes a perturbation to both phase and amplitude (mainly $|\psi_3|$). 
The red dot indicates the parameters simulated below. 
}
\label{State}
\end{figure}

An example of the ground state of a frustrated system is given in
Fig. \ref{State}.  Here 
$\eta_{13}=\eta_{23}<0$, is varied on the x-axis, while $\eta_{12}>0$ is kept constant. 
The resulting phases are shown in (B) and reveal a transition point at $\sim -0.0578$, that separates two regions with different phase-locking patterns. The state on the left spontaneously breaks $U(1)$ symmetry, but on the right side, time-reversal symmetry is also broken since the resulting state is not invariant under $\varphi_i\to -\varphi_i$. 
Broken time-reversal symmetry implies an additional twofold degeneracy of the ground state, and thus the transition is between the broken symmetries $U(1)$ and $U(1)\times Z_2$ respectively.  Increasing $\eta_{13}=\eta_{23}$ further, $Z_2$ symmetry is restored at $\sim -0.0542$, see (B) in Fig. \ref{State}.

A phase diagram of this type has been studied as a function of doping in connection with the iron based superconductor $Ba_{1- x}K_x Fe_2As_2$.
At the level of mean-field theory it features a $U(1)\times Z_2$ phase at low temperature (i.e. an $s+is$ state) \cite{maiti,nagaosa,tesanovic}.
The corresponding London model has been studied beyond the mean-field approximation, and it was then shown that fluctuations produce 
an additional phase where $Z_2$ symmetry is broken but $U(1)$ is restored \cite{PhysRevB.88.220511,PhysRevB.89.104509}.
The effect which we discuss
below revises these phase diagrams. 
This effect appears if density fluctuations are taken into account besides phase fluctuations.
\begin{figure*}[!htb]
 \hbox to \linewidth{ \hss
\includegraphics[width=\linewidth]{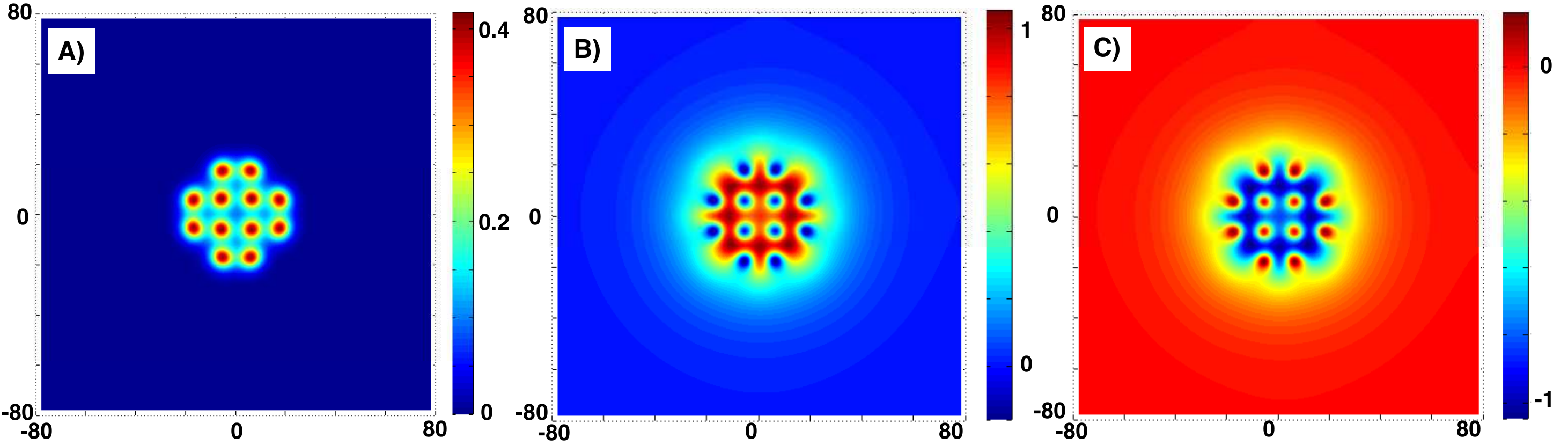}
 \hss}
\caption{
Vortex group solutions resulting from energy minimisation of the model (\ref{Hamiltonian}) on a two-dimensional grid. An initial configuration of 12 vortices was prepared and the energy minimisation was carried out with the constraint that the positions of the centers of the vortex cores remain fixed. 
This problem has two solutions that share the same magnetic field $\nabla\times \mathbf{A}$, shown in (A). 
However the solutions do not exhibit the same phase difference between the components.
The panels
(B) and (C) display $|\psi_1||\psi_2|\sin(\varphi_1-\varphi_2)$ for these two cases. 
The GL parameter set is given by the red dot in Fig. \ref{State} and features a ground state that breaks $U(1)$ symmetry only, implying that in the ground state $\sin(\varphi_1-\varphi_2)=0$. The presence of vortices however induces a phase difference between the components $\psi_1$ and $\psi_2$ which is positive in the first solution (B) and negative in the second solution (C). Away from the vortices, the phase differences decay exponentially to zero. The group thus produces a bubble of $Z_2$ order parameter.  
}
\label{cluster}
\end{figure*}

An important point here is that the interband terms take the form
\bea
\eta_{ab}|\psi_a||\psi_b|\cos (\varphi_a-\varphi_b), \label{interband}
\eea
i.e. they are modulated by the amplitudes. Since the phase-locking pattern is potentially altered by changing the parameters $\eta_{ij}$, it follows that perturbations to the amplitudes also have this capacity.  This gives vortices a very particular role in this model, since vortex cores suppress densities in a non-trivial way. 

The parameter set on which we focus here (marked by the red dot in Fig. \ref{State}) features broken $U(1)$ symmetry in the ground state. 
 It is clear from (B), that the system undergoes a symmetry change when the magnitude of $\eta_{13}=\eta_{23}$ is diminished. However according to Eq. \ref{interband}, this effect can likewise be obtained by depleting the amplitude $|\psi_3|$.

Vortex excitations in this three-band model are composite { (i.e. can be viewed as
a bound state of vortices with $2\pi$ winding in each of the phases $\varphi_{a}$)} and have three cores which, in general have different sizes
 (for a detailed study of the different characteristic length scales in vortex cores in such models see \cite{Johan2011,GL_TWO_BAND}) .
For the  parameters considered here, the third component has the largest vortex cores.  The consequence of this is that vortices deplete the three inter-band interaction terms to different extent, affecting interactions that involve the third component more. 
For a sufficiently dense group  of vortices, this results in the
formation of a bubble of induced non-trivial  phase difference between the superconducting components,
on average different from $0$ or $\pi$. That is,  a region of fluctuating $Z_2$ order parameter. 

An example of this is given in Fig. \ref{cluster}. 

The data in this figure was obtained by numerical minimization of the model (\ref{Hamiltonian}), which was carried out on a two-dimensional grid.
The system was prepared with a group of 12 numerically pinned vortices, and the energy was minimised subject to the constraint that the vortex-core positions remained unchanged. 
This problem has two solutions that share the same distribution and direction of magnetic flux (A). 
The plots (B,C) correspond to different solutions and 
reveal an induced phase difference between the components $\psi_1$ and $\psi_2$, which can be either positive (B) or negative (C). 
See also remark 
\footnote{ {Note that in principle, in this kind of models
the appearance of the bubbles of fluctuating $Z_2$ order parameter is not necessarily correlated with 
vortices. However, we are specifically interested in situations where there is a strong correlation of this type.
For the parameter set which we consider, small perturbations of the phase difference decouple from the density
fluctuations. Thus, the appearance of a fluctuating $Z_2$ order parameter requires strong density perturbations such as the presence of vortex cores which appear here 
as a consequence of fluctuations in the $U(1)$ sector of the model.
}}

We now turn to the main question of  this paper: whether topological excitations with this property drive a transition to a state with spontaneously broken  
 $Z_2$ symmetry upon heating.
Unless the systems is strongly Type-1, thermal fluctuations in the $U(1)$ sector results in excitation of vortex loops. 
These tend to disorder the $U(1)$ sector, but at the same time, as   shown above, they create bubbles of fluctuating $Z_2$ order parameter.
Indeed as long as the vortex loops are finite and well-separated this cannot lead to breakdown of $Z_2$ symmetry.
However a conjecture
which we investigate below is that once the density of vortex loops in the system grows to some characteristic value, the bubbles  with ``locally broken" $Z_2$ symmetry form a connected network that spans the entire system. This in turn can lead to spontaneously broken time-reversal symmetry in the system that results from heating. 
Restoration of the symmetry requires further heating to higher temperature. The phase with broken $Z_2$ symmetry thus exists between two characteristic temperatures.

To test this hypothesis we have conducted large scale Monte Carlo simulations of the model (\ref{Hamiltonian}) using the metropolis algorithm. 
In the discretised version of the Hamiltonian the covariant derivative and magnetic flux take the form
\bea
|\mathbf{D}_x\psi_{a,ijk}|^2 &=&  |\psi_{a,ijk}-\psi_{a,(i+1)jk} e^{ie \mathbf{A}_{xijk}}|^2,\\
\mathbf{B}_{zijk}&=&\mathbf{A}_x(i,j,k)+\mathbf{A}_y(i+1,j,k)\\
&-& \mathbf{A}_x(i,j+1,k)-\mathbf{A}_y(i,j,k)
\eea
where the subscript $xijk$ means vector component $x$ on the lattice point $ijk$ and so forth.
The discrete Hamiltonian is then given by
\bea
H= \sum_{i,j,k}\Big\{  \sum_{a=1}^3 \frac{1}{2}|\mathbf{D}\psi_{a,ijk}|^2+\frac{1}{2}\mathbf{B}_{ijk}^2+U_{ijk} \Big\}
\label{DHamiltonian}
\eea
where the last term is the potential which does not depend on gradients. 
 The corresponding partition function is given by 
$
Z=\int \mathcal{D}\mathbf{A}(\mathbf{r})  \mathcal{D}\psi_1(\mathbf{r}) \mathcal{D}\psi_2(\mathbf{r}) \mathcal{D}\psi_3(\mathbf{r}) e^{-\beta H},
$
with the inverse temperature $\beta$. 
{
The model \ref{Hamiltonian} is expressed in dimensionless units, with a length scale that is equal to the lattice spacing. }
The parameters used in the simulations are given in Fig. \ref{State} with $\eta_{13}=\eta_{23}=0.0611$. 
At zero temperature this system breaks $U(1)$ symmetry only. 
The figure also gives the masses of normal modes which, by definition are inverse coherence lengths. In this model they are associated with linear combinations of the fields $\psi_a$ \cite{Johan2011}.
The effects which we discuss do not require fine-tuning. Our   parameters
  are selected so that all the length scales are bigger than the lattice spacing but smaller than the system size.
\footnote{The simulations were conducted on cubic lattices with system sizes $20\le L\le 56$, periodic boundary conditions and a lattice spacing of $1$, meaning that at $T=0$, the shortest coherence length is almost twice the lattice spacing (with a mass of $\sim 0.6$).  
For each system size, simulations were conducted at 384 inverse temperatures which were uniformly distributed in the range $0.6\le \beta\le 2.5$. The large number of inverse temperatures allowed for parallel tempering to be employed. 
In all simulations, every data point was updated at least $1.5\times10^7$ times.}

To construct the order parameter associated with time-reversal symmetry breaking
we introduce a projection of the configuration space $\{\psi_1,\psi_2,\psi_3\}\to \pm1$ given by:   
$
f(\bar{\varphi})=\mathbf{sgn}\Big(\sin[\varphi_3](-\cos[\varphi_1]+\cos[\varphi_2])
+\sin[\varphi_1](-\cos[\varphi_2]+\cos[\varphi_3])
+\sin[\varphi_2](-\cos[\varphi_3]+\cos[\varphi_1])\Big),
$
which is odd under time reversal and changes sign if, and only if phases 
 are permuted.
Ordering in the $Z_2$ sector can then be determined by an order parameter that takes the same form as that of the Ising model:
\bea
O_{Z_2} =\Big\langle \Big|\sum_{k,l,m} f(\bar{\varphi}_{k,l,m})\Big|\Big\rangle\frac{1}{L^3}.\label{OZ2}
\eea
Restoration of the local $U(1)$ symmetry and thus the onset of the non-superconducting state can be identified by the scaling properties of the Fourier components of the magnetic field. We start by introducing
\bea
c=2L^{-3}\sum_{ijk}  \mathbf{B}_y \cos\frac{2\pi i}{L},\;
s=2L^{-3}\sum_{ijk}  \mathbf{B}_y \sin\frac{2\pi i}{L}.\;
\eea
In the normal state, the gauge field is massless and the expectation value of $c,s$ is given by
\bea
\langle s^2\rangle=\langle c^2\rangle = \frac{\int dc \; c^2 e^{-\beta L^3 c^2/4}}{\int dc \;  e^{-\beta L^3 c^2/4}}=\frac{2}{\beta L^3}.
\eea
We thus define 
\bea
F_A(L,\beta)=L^{3} \langle c^2+s^2\rangle \label{FA}
\eea
which should be scale invariant in the non-superconducting state. Plotting $F_A(L,\beta)$ versus $\beta$ for several system sizes, we expect the curves to collapse onto the same line once $U(1)$ symmetry is restored. 

To determine how thermally excited vortex loops affect the $O_{Z_2}$ order parameter we define the total length of all vortex lines ${\varrho}$, and introduce the quantity
\bea
\rho_V={ \varrho}{L^{-3}}\label{rhov},
\eea
which allows us to define the correlator of the amount of thermally induced vortex matter and the order parameter $O_{Z_2}$: 
\bea
C_{V,Z_2}=\frac{\langle  \rho_V O_{Z_2} \rangle-\langle  \rho_V  \rangle \langle   O_{Z_2} \rangle }{\sigma(\rho_V)\sigma(O_{Z_2})},\label{CVZ2}
\eea
where $\sigma$ denotes the standard deviation.

\begin{figure*}[!htb]
 \hbox to \linewidth{ \hss
\includegraphics[width=\linewidth]{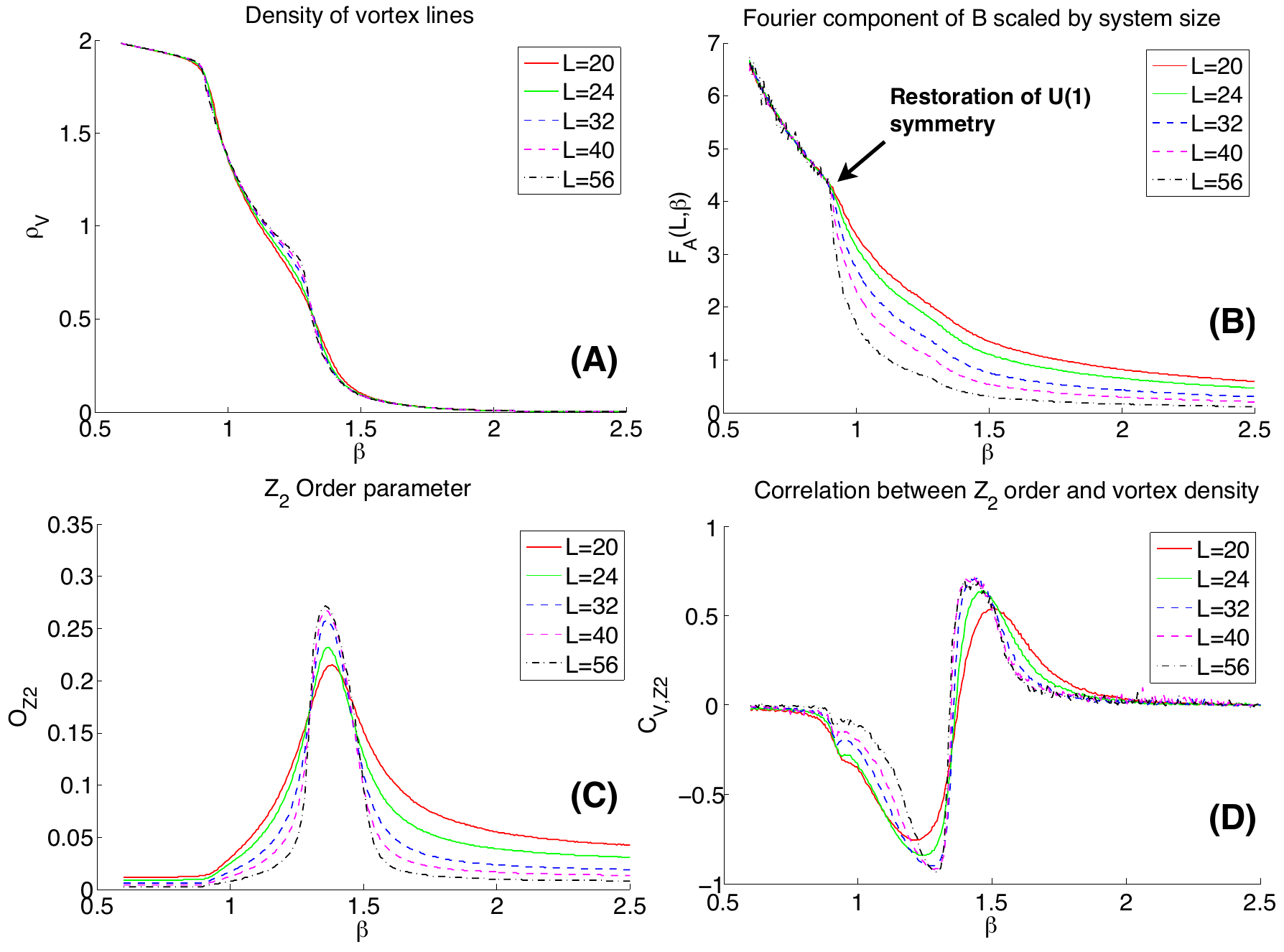}
 \hss}
\caption{
Summary of the simulation results. 
(A) The density of thermally induced vortices ($\rho_V$, Eq. \ref{rhov}). It approaches zero at low temperature, but starts to increase substantially at $\beta\approx 1.5$. 
 (B) The Fourier components of the magnetic field scaled by system size ($F_A(L,K)$, Eq. \ref{FA}) 
 collapse onto the same curve, indicating that the system is non-superconducting for $\beta<\beta_c\approx 0.9$. 
(C) The order parameter of the broken time-reversal symmetry $O_{Z_2}$ (Eq. \ref{OZ2})
is zero at low temperature (except for finite size effects). As the temperature increases an order emerges and reaches a maximum at $\beta\approx 1.35-1.4$. At higher temperatures this order starts to decay rapidly.
(D) The correlation between the density of thermally-induced vortices $\rho_V$ and the Ising order parameter $O_{Z2}$ (Eq. \ref{CVZ2}) is positive at low temperature and reaches a maximum of $\sim 0.72$, implying that breaking of the $Z_2$ symmetry is driven by vortex proliferation.  
At higher temperatures the correlation becomes negative, indicating that vortices contribute to restoring the symmetry. 
}
\label{results}
\end{figure*}

The results of the simulations, shown in Fig. \ref{results} confirm the scenario described above. 
At low temperature the system does not break time-reversal symmetry. Note that in that case $O_{Z_2}$ is only nonzero due to finite size effects: it decreases rapidly with system size. As the temperature increases,  we see the onset of a genuine $Z_2$ order, which reaches a maximum at $\beta\approx 1.35-1.4$. 
This symmetry change is primarily driven by excitation of vortices (as opposed to non-topological fluctuations). This is clear from the correlator $C_{V,Z_2}$ shown in (D), which reaches $\sim 0.72$, indicating a very strong correlation between the density of vortices and the order parameter $O_{Z_2}$. It is also consistent with the fact that the $Z_2$ order parameter is only nonzero in a temperature region where there is an appreciable density of vortices.

As the temperature increases further, $O_{Z_2}$ starts to decrease as expected. While  thermal fluctuations generally tend to restore broken symmetries, an additional effect is also present. 
At higher temperatures the correlator $C_{V,Z_2}$ becomes strongly negative, suggesting that further increase in the density of thermally induced vortices helps to destroy the $Z_2$ order. Returning to Fig. \ref{State} (B), it  is clear that in the model we use, the region with broken time-reversal symmetry corresponds to {\it intermediate} magnitudes of $\eta_{13}=\eta_{23}$. Decreasing the magnitude beyond $|\eta_{i3}|\sim 0.0542$ restores time-reversal symmetry and results in the ``$s_{\pm}$ state" which is characterized by phase ``anti-locking", i.e. $\varphi_1-\varphi_2=\pi$. Likewise, depleting $|\psi_3|$ beyond a certain point contributes to destroying the $Z_2$ order by the same mechanism. 
The other process which should, in general, contribute to the anticorrelation at elevated temperatures is the
 splitting of composite vortices into fractional ones connected by $Z_2$ domain walls (for a detailed
 discussion of these objects see \cite{PhysRevLett.107.197001,chiralcp2}).

In conclusion, it is well known that  broken symmetries can be restored by entropy-driven proliferation of topological defects. Here we have shown that for a class of systems, the proliferation of topological defects  instead leads to a spontaneous breakdown of an additional symmetry. 
The implication of this is a phase transition where a symmetry is broken as the temperature is increased.
We have demonstrated this effect using a three-component GL model with frustrated interband interaction as an example. These models are currently discussed in connection $Ba_{1-x}K_{x}Fe_2As_2$ for a certain range of dopings.  This transitions should be detectable in calorimetry experiments.
  The mechanism described here is however more generic and should also apply to other systems where 
topological defects induce a bubble of fluctuating order parameter associated with a different symmetry.

This work was supported
by the Knut and Alice Wallenberg Foundation through a Royal Swedish
Academy of Sciences Fellowship, by the Swedish Research Council grants 
642-2013-7837,  325-2009-7664, and. Part of the work was done at University of Massachusetts Amherst supported by  the NSF CAREER Award DMR-0955902.
{Computations were performed on resources provided by the Swedish National Infrastructure
for Computing (SNIC)  at the National Supercomputer Center in Link\"oping, Sweden.}

\bibliography{biblio}

\end{document}